\documentstyle[12pt]{article}

\def\susyb{{\begin{picture}(25,0)(0,0)
\put(0,0){\scriptsize $SUSY$}
\put(0,0){\line(4,1){22}}
\end{picture}}}
\def\Tr{\mbox{Tr}}
\def\A{{\cal A}}
\def\B{{\cal B}}

\catcode`\@=11
\long\def\@makefntext#1{
\protect\noindent \hbox to 3.2pt {\hskip-.9pt
$^{{\ninerm\@thefnmark}}$\hfil}#1\hfill}		

\def\@makefnmark{\hbox to 0pt{$^{\@thefnmark}$\hss}}  

\def\ps@myheadings{\let\@mkboth\@gobbletwo
\def\@oddhead{\hbox{}
\rightmark\hfil\ninerm\thepage}
\def\@oddfoot{}\def\@evenhead{\ninerm\thepage\hfil
\leftmark\hbox{}}\def\@evenfoot{}
\def\sectionmark##1{}\def\subsectionmark##1{}}

\renewcommand{\thefootnote}{\fnsymbol{footnote}}

\newcounter{sectionc}\newcounter{subsectionc}\newcounter{subsubsectionc}
\renewcommand{\section}[1] {\vspace*{0.6cm}\addtocounter{sectionc}{1}
\setcounter{subsectionc}{0}\setcounter{subsubsectionc}{0}\noindent
	{\normalsize\bf\thesectionc. #1}\par\vspace*{0.4cm}}
\renewcommand{\subsection}[1] {\vspace*{0.6cm}\addtocounter{subsectionc}{1}
	\setcounter{subsubsectionc}{0}\noindent
	{\normalsize\it\thesectionc.\thesubsectionc. #1}\par\vspace*{0.4cm}}
\renewcommand{\subsubsection}[1]
{\vspace*{0.6cm}\addtocounter{subsubsectionc}{1}
	\noindent {\normalsize\rm\thesectionc.\thesubsectionc.\thesubsubsectionc.
	#1}\par\vspace*{0.4cm}}

\newcounter{appendixc}
\newcounter{subappendixc}[appendixc]
\newcounter{subsubappendixc}[subappendixc]

\renewcommand{\appendix}[1] {\vspace*{0.6cm}
        \refstepcounter{appendixc}
        \setcounter{figure}{0}
        \setcounter{table}{0}
        \setcounter{equation}{0}
        \renewcommand{\thefigure}{\Alph{appendixc}.\arabic{figure}}
        \renewcommand{\thetable}{\Alph{appendixc}.\arabic{table}}
        \renewcommand{\theappendixc}{\Alph{appendixc}}
        \renewcommand{\theequation}{\Alph{appendixc}.\arabic{equation}}
        \noindent{\bf Appendix \theappendixc #1}\par\vspace*{0.4cm}}

\def\abstracts#1{{
	\centering{\begin{minipage}{12.2truecm}\normalsize\baselineskip=12pt\noindent
	\centerline{\normalsize ABSTRACT}\vspace*{0.3cm}
	\parindent=0pt #1
	\end{minipage}}\par}}

\renewenvironment{thebibliography}[1]
	{\begin{list}{\arabic{enumi}.}
	{\usecounter{enumi}\setlength{\parsep}{0pt}
\setlength{\leftmargin 1.25cm}{\rightmargin 0pt}
	 \setlength{\itemsep}{0pt} \settowidth
	{\labelwidth}{#1.}\sloppy}}{\end{list}}

\newcounter{itemlistc}
\newcounter{romanlistc}
\newcounter{alphlistc}
\newcounter{arabiclistc}

\newcommand{\fcaption}[1]{
        \refstepcounter{figure}
        \setbox\@tempboxa = \hbox{\footnotesize Fig.~\thefigure. #1}
        \ifdim \wd\@tempboxa > 6in
           {\begin{center}
        \parbox{6in}{\footnotesize\baselineskip=12pt Fig.~\thefigure. #1}
            \end{center}}
        \else
             {\begin{center}
             {\footnotesize Fig.~\thefigure. #1}
              \end{center}}
        \fi}

\newcommand{\tcaption}[1]{
        \refstepcounter{table}
        \setbox\@tempboxa = \hbox{\footnotesize Table~\thetable. #1}
        \ifdim \wd\@tempboxa > 6in
           {\begin{center}
        \parbox{6in}{\footnotesize\baselineskip=12pt Table~\thetable. #1}
            \end{center}}
        \else
             {\begin{center}
             {\footnotesize Table~\thetable. #1}
              \end{center}}
        \fi}

\def\@citex[#1]#2{\if@filesw\immediate\write\@auxout
	{\string\citation{#2}}\fi
\def\@citea{}\@cite{\@for\@citeb:=#2\do
	{\@citea\def\@citea{,}\@ifundefined
	{b@\@citeb}{{\bf ?}\@warning
	{Citation `\@citeb' on page \thepage \space undefined}}
	{\csname b@\@citeb\endcsname}}}{#1}}

\newif\if@cghi
\def\cite{\@cghitrue\@ifnextchar [{\@tempswatrue
	\@citex}{\@tempswafalse\@citex[]}}
\def\citelow{\@cghifalse\@ifnextchar [{\@tempswatrue
	\@citex}{\@tempswafalse\@citex[]}}
\def\@cite#1#2{{$\null^{#1}$\if@tempswa\typeout
	{IJCGA warning: optional citation argument
	ignored: `#2'} \fi}}

 1
 1
 1

\font\ninerm=cmr9

\begin{document}

\newcommand{\st}{\scriptstyle}
\newcommand{\sst}{\scriptscriptstyle}
\newcommand{\mco}{\multicolumn}
\newcommand{\epp}{\epsilon^{\prime}}
\newcommand{\vep}{\varepsilon}
\newcommand{\ra}{\rightarrow}
\newcommand{\ppg}{\pi^+\pi^-\gamma}
\newcommand{\vp}{{\bf p}}
\newcommand{\ko}{K^0}
\newcommand{\kb}{\bar{K^0}}
\newcommand{\al}{\alpha}
\newcommand{\ab}{\bar{\alpha}}
\def\be{\begin{equation}}
\def\ee{\end{equation}}
\def\bea{\begin{eqnarray}}
\def\eea{\end{eqnarray}}
\def\CPbar{\hbox{{\rm CP}\hskip-1.80em{/}}}

\renewcommand{\thefootnote}{\fnsymbol{footnote}}

\begin{center}
Mar 15, 1995 \hfill LBL-36962
\end{center}

\vfill
\centerline{\large\bf NON-UNIVERSAL SUSY BREAKING,}
\baselineskip=22pt
\centerline{\large\bf HIERARCHY AND SQUARK
DEGENERACY\footnotemark~\footnotemark}

\addtocounter{footnote}{-1}
\footnotetext{This
work was supported by the Director, Office of Energy Research,
Office of High Energy and Nuclear Physics, Division of High
Energy Physics of the U.S. Department of Energy
under Contract DE-AC03-76SF00098.}
\stepcounter{footnote}
\footnotetext{Invited plenary talk presented at ``Beyond the Standard
Model IV,'' Dec 13 to Dec 18, 1994, Granlibakken, Lake Tahoe, California.}

\vfill
\vspace*{0.6cm}
\centerline{\normalsize HITOSHI MURAYAMA\footnotemark}
\baselineskip=13pt
\centerline{\normalsize\it Theoretical Physics Group, Lawrence
Berkeley Laboratory}
\baselineskip=12pt
\centerline{\normalsize\it University of California, Berkeley, CA 94720}
\centerline{\normalsize E-mail: murayama@lbl.gov}

\footnotetext{On leave of absence
from \it Department of Physics, Tohoku University, Sendai, 980 Japan.}

\vfill
\vspace*{0.9cm}
\abstracts{I discuss non-trivial effects in the soft SUSY breaking terms
which appear when
one integrates out heavy fields.  The effects exist only when the SUSY
breaking terms are non-universal.  They may spoil (1) the hierarchy
between the weak and high-energy scales, or (2) degeneracy among the
squark masses even in the presense of a horizontal symmetry. I argue, in
the end, that such new effects may be useful in probing physics at
high-energy scales from TeV-scale experiments. }

\vfill
\newpage

\mbox{ }

\vskip 1in

\begin{center}
{\bf Disclaimer}
\end{center}

\vskip .2in

\begin{scriptsize}
\begin{quotation}
This document was prepared as an account of work sponsored by the United
States Government. While this document is believed to contain correct
information, neither the United States Government nor any agency
thereof, nor The Regents of the University of California, nor any of their
employees, makes any warranty, express or implied, or assumes any legal
liability or responsibility for the accuracy, completeness, or usefulness
of any information, apparatus, product, or process disclosed, or represents
that its use would not infringe privately owned rights.  Reference herein
to any specific commercial products process, or service by its trade name,
trademark, manufacturer, or otherwise, does not necessarily constitute or
imply its endorsement, recommendation, or favoring by the United States
Government or any agency thereof, or The Regents of the University of
California.  The views and opinions of authors expressed herein do not
necessarily state or reflect those of the United States Government or any
agency thereof or The Regents of the University of California and shall
not be used for advertising or product endorsement purposes.
\end{quotation}
\end{scriptsize}

\vskip 2in

\begin{center}
\begin{small}
{\it Lawrence Berkeley Laboratory is an equal opportunity employer.}
\end{small}
\end{center}

\newpage
\setcounter{page}{1}
\renewcommand{\thefootnote}{\alph{footnote}}
\setcounter{footnote}{0}

\normalsize\baselineskip=15pt
\setcounter{footnote}{0}
\renewcommand{\thefootnote}{\alph{footnote}}

\section{Introduction}

The Standard Model (SM) of the particle physics is extremely succesfull
and is now being tested experimentally at a precision better than 1~\%
level.  However, it leaves many questions unanswered: the origin of
flavor, rather complex quantum numbers under the gauge group, anomaly
cancellation, charge quantization, and many others.  Any attemps to
build models which answer these questions involve new physics at much
deeper levels, {\it e.g.}\/, much higher energies.  Then one has to
ensure that the hierarchy between the weak scale and the energy
scales of new physics is stable under the radiative corrections.
Supersymmetry (SUSY) has been regarded as a promising candidate to
ensure the stability of such a hierarchy.

SUSY, however, has also many problems especially from the model
building point of view.  First of all, there is no concensus how the
supersymmetry is broken.  It tends to give too large rates for the
flavor-changing neutral current processes.  And, the most importantly,
supersymmetry itself does not explain the hierarchy; it merely
stabilizes it.  For a more complete list of the problems, I refer to
a talk by Haber.\cite{top-ten}

In this talk, I point out several other problems in SUSY model building
which, to my understanding, are not widely recognized; these problems
arise only when the SUSY breaking terms are non-universal.  The first is
that the hierarchy may be spoiled by the SUSY breaking effect.  The
second is that the degeneracy among the scalar quarks may not be
guaranteed even with the horizontal symmetries.  Both of the problems can
be discussed within the same context: integrating out heavy fields in
the presence of the SUSY breaking effects.  Integrating out the heavy
fields is not the same as throwing them away; they leave non-trivial
relics in the soft SUSY breaking terms in the low-energy effective
theory.  I will exemplify how non-trivial effects arise in the next
few sections.

Let me remind you that having many heavy fields at a mass scale $M$
below the Planck scale $M_P$ is a relatively generic feature of the SUSY
models.  SUSY GUT of
course have many heavy fields at the GUT-scale $M \simeq 10^{16}$~GeV, and
they have to be integrated out.  Most of the flavor models also have
many heavy fields below the Planck scale; one uses $M/M_P \sim
0.01$--$0.1$ as a small expansion parameter to reproduce the
hierarchical structure of the Yukawa matrices.  Therefore, it is a
very general question to analyze the soft SUSY breaking terms when you
integrate out heavy fields.

\section{Naive Integration of Heavy Fields}

Let me first explain what kind of misconception I myself had in the
past.\footnote{All the discussion applies only to the framework where
the SUSY-breaking masses are fed into the fields of our interest at a
scale above the scale of the heavy fields which we integrate out,
{\it e.g.}\/, hidden sector models.  If the SUSY breaking effects appear
at very low-energy,\cite{Dine} the problems may not exist.}

Suppose we have a SUSY model with superheavy fields, {\it e.g.}\/, at
the scale of the grand unified theory (GUT).  When we derive the Minimal
Supersymmetric Standard Model (MSSM) from a typical SUSY GUT, we have to
integrate out the superheavy fields to obtain the MSSM as an effective
low-energy theory.  Of course the superpotential of the model has to be
chosen such that the doublet Higgs superfields in the MSSM have masses
only of $O(m_W)$, either by a fine-tuning or some other ``natural''
mechanisms.  {\sl So far it is completely true.}

When we integrate out the heavy fields, the SUSY breaking effects
are negligible, since they are much smaller than the physics scale under
discussions, {\it e.g.}\/, $m_{SUSY} \ll M_{GUT}$.  Therefore, we can
integrate out the heavy fields without the SUSY breaking effects in
mind, and write down the SUSY Lagrangian of the MSSM.  Then we introduce
SUSY breaking terms later, at $O(m_{SUSY})$.  The SUSY breaking terms in
the MSSM satisfy boundary conditions dictated by the symmetries of the
original theory, such as GUT symmetry or horizontal symmetries.
For instance, an $SU(2)$ horizontal symmetry
between the first and second generations
guarantees $m_{\tilde{d}} = m_{\tilde{s}}$.  {\sl This is completely
wrong.}

There are two main mistakes in the results we obtain under this
``naive'' integration of the heavy fields.  First, the ``light'' fields
in the low-energy theory may have SUSY-breaking mass terms which are
much bigger than $O(m_{SUSY}^2)$, thereby spoiling the hierarchy.
Second, the SUSY-breaking masses in the low-energy theory may not
respect the symmetries in the original theory at all.  These are the
points which I'll explain in this talk.  Although these cases are
problematic, it is welcome in general to have effects of the heavy
fields in the soft SUSY breaking terms of the light fields.  What we
learn here is that the SUSY-breaking masses are much more sensitive to
the physics at very high energy scales than we naively think.  This
opens up a wider ``window'' to the physics at very high energy scales
for us.\cite{ICEPP}

\section{General Soft SUSY Breaking Terms}

Under the popular assumption of the ``minimal supergravity'' (or
``universal'' SUSY breaking terms), the soft SUSY breaking terms take the
following form:
\begin{equation}
V_{\susyb} = \A W + \B z^i F^*_i + h.c.,
\end{equation}
where $W$ is the superpotential, $F^i$ is the auxiliary component of the
chiral supermultiplet whose scalar component is $z^i$, and $\A$, $\B$ are
dimensionful parameters of $O(m_{SUSY})$. This form may look
unfamliar, but it should look familiar after integrating out the
auxiliary fields:
\begin{equation}
V_{\susyb} = (\A+3\B) W_3 + (\A+2\B) W_2 + (\A+\B) W_1 + h.c. + \B^2 |z^i|^2 .
\end{equation}
Here, $W_3$ contains trilinear terms in the superpotential $W$, $W_2$
bilinear, and $W_1$ linear.  Actually, one can prove that the ``naive''
integration of the heavy fields explained in the previous section is
exact up to a redefinition of the $\A$ and $\B$ parameters; this amazing
result was derived by Hall, Lykken and Weinberg\cite{HLW} a decade ago.

The most general soft SUSY breaking terms can be written as follows if
one does not assume the ``universality,''
\begin{equation}
V_{\susyb} = \A_m W_m + \B_i^j z^i F^*_j + h.c. + (m^2)_i^j z^i z^*_j ,
\end{equation}
where $W_m$ refers to invidual terms in the superpotential with
arbitrary independent SUSY breaking coefficients $\A_m$.  The parameter $\B$ in
the universal case is extended to be an arbitrary matrix in the field
space.  In addition, one can add arbitrary scalar mass matrix $m^2$.

There are at least three reason why we want to consider non-universal
SUSY breaking terms at the scale $M$ where we integrate out heavy fields.  (1)
They may be non-universal already at $M_P$, like in superstring
theories.  (2) Universal SUSY breaking terms are not stable under the
renormalization, and hence may be corrected by the physics at the Planck
scale.  (3) Their running from $M_P$ to $M$ spoil the universality.
Therefore, we have to integrate out heavy fields in the presence of
non-universal SUSY breaking terms.

\section{Spoiling Hiearchy by SUSY-breaking Effects}

In this section I present two examples where the fields which have only
$O(m_{SUSY})$ masses in the superpotential can acquire soft SUSY breaking
masses of $O(m_{SUSY} M_X)$, where $M_X$ is the scale of the heavy fields
you are integrating out.

The first one is the famous example of the minimal $SU(5)$ GUT by
Dimopoulos, Georgi\cite{DG} and Sakai.\cite{Sakai}  The $SU(5)$
symmetry is broken by an adjoint Higgs superfield $\Sigma$, and the
Higgs doublets belong to $SU(5)$ quintets, $H_u$ and $H_d$.  The
superpotential of this model is
\begin{equation}
W = \lambda {\rm tr} \Sigma^3 + M_\Sigma {\rm tr} \Sigma^2
	+ H_u (f \Sigma + M_H) H_d.
\end{equation}
$\lambda$, $f$ are
dimensionless coupling constants, while $M_\Sigma$, $M_H$ are GUT scale
mass parameters.
We add the most general SUSY breaking terms,
\begin{eqnarray}
V_\susyb = \lambda A_\Sigma {\rm tr} \Sigma^3
		+ B_\Sigma M_\Sigma {\rm tr} \Sigma^2
		+ f A_H H_u \Sigma H_d + B_H M_H H_u H_d + O(m_{SUSY}^2),
\end{eqnarray}
where $A_\Sigma$, $B_\Sigma$, $A_H$ and $B_H$ are the SUSY breaking
parameters of order $m_{SUSY}$. Taking
$\Sigma = \mbox{diag}(2,2,2,-3,-3) \sigma$,
the minimum of the potential lies at $
\sigma_0 = 2 M_\Sigma/3\lambda $ in the SUSY limit, which is
shifted by $\delta \sigma=
(A_{\Sigma}-B_{\Sigma})/3\lambda$ in the presence of the SUSY breaking terms.
The mixing mass of
the two doublet Higgs bosons\footnote{Hereafter
$H_u$ and $H_{d}$ represent the $SU(2)_L$ doublet Higgs multiplets.}\/
$m_{12}^2 H_u H_d$ is given by
\begin{equation}
m_{12}^2 =  3f \sigma_0
		(A_\Sigma - B_\Sigma - A_H + B_H)+O(m_{SUSY}^2),
\end{equation}
where we have used that the supersymmetric mass of the  Higgs
doublets is fine-tuned to be
$M_{H}-3f\sigma_0 =O(m_{SUSY})$
in the superpotential.  Clearly for a class of the SUSY breaking
parameters where the combination $A_\Sigma - B_\Sigma - A_H + B_H $ does
not vanish,
$m_{12}^2$ lies at an intermediate scale $\sim m_{SUSY} M_X$ and the
gauge hierarchy is spoiled.

One may anticipate that such a problem exists only for models which have
fine-tunings as this example.  I would argue, however, that this
problem is rather generic.  For instance, such a problem may arise even
without a GUT symmetry.  Let us denote doublet Higgs fields in the MSSM
by $H_u$ and $H_d$.  Suppose there is some reason that
no mass term exists for $H_u$ and $H_d$ in the superpotetial in
the absence of SUSY breaking, and also that SUSY is broken in the hidden
sector by a O'Raifeartaigh sector for definiteness.  Then there is a
chiral superfield $X$ in the hidden sector which has a vacuum
expectation value in the $F$-component, $\langle X \rangle \sim \theta^2
m_{SUSY} M_P$. Since we have to generate $\mu$-term anyway, we need a
coupling as $\int d^4\theta X^* H \overline{H}/M_P$.  But then we could
also have a coupling $\int d^2 X H \overline{H}$, which again leads to a
too-large soft SUSY breaking mass term to the Higgs bosons.

Actually, one can prove that such a problem does not occur in a slightly
restricted form of the SUSY breaking terms,\cite{KMY2}
\begin{equation}
V_\susyb = \A W + \B_i^j z^i
    \frac{\partial W}{\partial z^j}+h.c.
	+ O(m_{SUSY}^2) \ {\rm terms}.
\label{special}
\end{equation}
In particular, one automatically obtains the relation $A_\Sigma -
B_\Sigma - A_H + B_H = 0$ in the minimal $SU(5)$ from this ansatz with
no other additional constraints.  This ansatz for the SUSY breaking
terms have two nice features: (1) non-universal enough such that the
form is stable under renormalization, and (2) still restricted enough to
guarantee the hierarchy.  Indeed, one can derive a general formula for
the soft SUSY breaking terms after integrating out heavy
fields.\cite{KMY2}

\section{How Squark Degeneracy May Be Spoiled}

In this section, I present a toy model with a global horizontal $SU(2)$
symmetry.  Even though the $SU(2)$ symmetry was meant to guarantee the
degeneracy between the first- and second-generation squarks, it actually
doesn't in this example.

Let me first explain how additional contribution ($F$-term
contribution)\cite{KMY2} can be generated to the scalar mass term in the
low-energy effective theory in general by integrating out a heavy field.
Suppose there is a vector-like heavy fields $\psi$ and $\overline{\psi}$
with a mass term $M\overline{\psi} \psi$, and a light chiral field
$\phi$ which does not have a mass term.  However, the heavy and light
fields mix by picking up a vacuum expectation value of the field $\chi$.
The superpotential is
\begin{equation}
W = M \overline{\psi} \psi + g \overline{\psi} \chi \phi
	+ \lambda \left( \frac{1}{3} \chi^3 - \frac{1}{2} v \chi^2 \right).
\end{equation}
The vacuum is $\langle \chi \rangle = v$.  The heavy and light fields
$\psi'$ and $\phi'$ are defined by
\begin{eqnarray}
& & \psi' = \frac{1}{\sqrt{M^2 + (gv)^2}} (M \psi + gv \phi),\\
& & \phi' = \frac{1}{\sqrt{M^2 + (gv)^2}} (-gv \psi + M \phi).
\end{eqnarray}
Now the SUSY breaking terms have tri-linear and bi-linear terms as well as the
scalar mass terms
\begin{equation}
V_\susyb = B_\psi M \overline{\psi} \psi + g A_\phi \overline{\psi} \chi \phi
	+ \lambda \left(  \frac{A_\chi}{3} \chi^3
		-  \frac{B_\chi}{2} v \chi^2 \right)
	+ m_\psi^2 |\psi|^2 + m_{\overline{\psi}}^2 |\overline{\psi}|^2
	+ m_\phi^2 |\phi|^2 + m_\chi^2 |\chi|^2.
\end{equation}
The point is that the three different mass matrices, namely
supersymmetric mass terms in the superpotential,
the SUSY breaking tri-linear and bi-linear terms, and SUSY breaking
scalar mass terms cannot be simultaneously
diagonalized if $A_\phi \neq A_\chi$ or $B_\psi \neq B_\chi$ or
$m_\phi^2 \neq m_\psi^2$.  The resulting scalar mass for $\phi'$
after integrating out heavy fields $\psi'$, $\overline{\psi}$ and $\chi$
is
\begin{equation}
m_{\phi'}^2 = \frac{M^2}{M^2 + (gv)^2} m_\phi^2
	+ \frac{(gv)^2}{M^2 + (gv)^2} m_\psi^2
	- \frac{M^2 (gv)^2}{[M^2 + (gv)^2]^2}
		(A_\chi - B_\chi + B_\psi - A_\phi)^2.
\end{equation}
If all the SUSY breaking terms are universal, {\it i.e.}\/ $A_\chi = A_\phi =
A$, $B_\chi = B_\psi = B$, and $m_\phi^2 = m_\psi^2 = m_0^2$, it
drastically simplifies to give $m_0^2$, and one can pretend nothing happened
by integrating out heavy fields.

Now comes the toy model with a horizontal $SU(2)_H$ symmetry.  Take
$\psi$, $\overline{\psi}$ and $\phi$ as left-handed quark fields
for the first two generations, each $SU(2)_H$ doublets.  We introduce
right-handed fields $D = (d_R^c, s_R^c)^T$ and $U = (u_R^c, c_R^c)^T$ as
well, both $SU(2)_H$ doublets.  We regard $\chi$
as $2\times 2$ matrix which breaks $SU(2)_H$ symmetry down to nothing
by its expectation value, $\langle \chi \rangle
= \mbox{diag}(v, V)$ with $v \ll V$.\footnote{One can write the most general
superpotential $W
= \mu^2 \Tr \chi + M_\chi \Tr \chi^2 + \lambda \Tr \chi^3$, with $\mu \ll
M$ and $\lambda \sim O(1)$.  Then this superpotential has a vacuum with
$v \sim \mu^2/M_\chi$ and $V \sim M_\chi$. $\chi$ is actually a reducible
representation under $SU(2)_H$, since it contains both singlet and
adjoint components.} Assume the following superpotential
\begin{equation}
W = M \overline{\psi} \psi +
	g_1 \overline{\psi} \chi \phi + h_s \psi D H ,
\end{equation}
with $M \sim V$.  It generates hierarchical Yukawa coupling constants
$h_d \sim h_s (v/M)$ after integrating out $\psi'$ and
$\overline{\psi}$.  Before integrating out the heavy fields, the
$SU(2)_H$ symmetry guarantees the same SUSY breaking masses for the both
components $\tilde{d}_L$ and $\tilde{s}_L$ in $\phi$.  However they have
different masses in the
low-energy effective theory because the mass $m_{\phi'}^2$ depends
on the $\langle \chi \rangle$.  Therefore the squark degeneracy is
broken even in the presence of a horizontal symmetry.  Similarly, you
also obtain similar contributions to the $\phi'$ mass from the up
sector.  And they do not commute each other, because of the Cabbibo
rotation.  Therefore not only you spoil the degeneracy between the
first- and second-generations, but also generate off-diagonal terms such
that quark and squark masses are not aligned.  Such a model
gives large rates for the flavor-changing neutral current processes
if not dead.

The situation is even worse when the horizontal symmetry is gauged.  As
known in linterature,\cite{D-term} there are additional
contributions to the scalar masses when the rank of the gauge group is
reduced by a symmetry breaking.  The easiest example of this phenomenon
is when $U(1)$ gauge symmetry is broken by charge $+1$ ($-1$) supefields
$H_+$ ($H_-$).  The expectation values of $H_+$ and $H_-$ can differ if
their SUSY breaking masses are different, $m_+^2 \neq m_-^2$:
\begin{equation}
\langle D \rangle =
\langle |H_+|^2 - |H_-|^2 \rangle = \frac{1}{g} (m_+^2 - m_-^2) ,
\end{equation}
where $g$ is the gauge coupling constant.  Then this condensation of the
$D$-component gives contributions to the masses of light scalar fields
$\phi^i$,
\begin{equation}
V = \sum_i g \langle D \rangle Q_i |\phi^i|^2
	= (m_+^2 - m_-^2) Q_i |\phi^i|^2 .
\end{equation}
Here $Q_i$ are the $U(1)$ charges of the fields $\phi^i$.
Note that the final result does not depend on the gauge coupling
constant; therefore one can never turn off the $D$-term contribution by
taking the gauge coupling constant arbitrarily small.\footnote{In the
previous example with a global horizontal symmetry, one could suppress
the additional contributions by taking the limit $M \rightarrow \infty$.
Of course there is a certain upper bound from the requirement that the
Yukawa coupling constants are not too small.  Then the question
becomes a numerical one.}  The gauged
horizontal symmetries give $D$-term contributions to the scalar masses
differently to the different generations, since they have different
quantum numbers under the horizontal symmetries.

\section{Final Remarks}

As we have seen, the integration of heavy fields leaves rather
non-trivial relics to the soft SUSY breaking term in the low-energy
effective theory.  They could be harmful in some cases: (1) it may spoil
the hierarchy, or (2) it may spoil the squark degeneracy.  Even though
these two cases are problematic, I would argue that it is actually
welcome to have non-trivial consequence of heavy fields in the
low-energy effective theory.  Of course, these effects put new
challenges to the model builders.  However, this also means that the
soft SUSY breaking terms in the low-energy effective theory are much
more sensitive to the physics at very high energy scale than we naively
think.  They have much richer structure than the ``universal'' case, at
least.  Therefore the future measurements of the soft SUSY breaking
parameters may allow us to figure out the symmetry structure at high
energies, flavor physics and so on.

Recall that the GUT-relation of the gaugino masses are very good
predictions of SUSY GUT.  The threshold corrections at the GUT-scale do
not generate large logarithms,\cite{HMG} and hence only of
$O(\alpha/\pi)$, as far as the gaugino masses are comparable to other
soft SUSY breaking terms from the beginning.  They satisfy the same
relation even in the presence of intermediate symmetries.\cite{KMY1}
Therefore GUT-relation of the gaugino masses provide us an excellent
tool to test the idea of SUSY GUT, and its test is experimentally
feasible.\cite{Tsukamoto}

Scalar masses are more sensitive to the detail of the physics at high
energies.\cite{KMY1} Even though the $F$-term contributions can in
general spoil the boundary conditions of GUT symmetry\footnote{This
point was recently emphasized by Dimopoulos and Pomarol,\cite{DP} and
their results are consistent with the general formula.\cite{KMY2}}, we
expect such effects are small enough for the first two generations to
suppress the flavor-changing effects adequately. Then the mass
measurements of squarks and sleptons of the first two generations at
future colliders can be still
used to test the symmetries at high energy scales.\cite{KMY1}  The
masses of the third generation fields and the Higgs bosons contain more
information on
the physics at high scales.  Finally, rare flavor-changing
effects,\cite{BH} CP-violating effects\cite{DH} and proton
decay\cite{AHMR} provide us probes to the tiny effects in the scalar
mass matrices from the flavor physics at high scales.  If we are lucky
enough to see many different kinds of signatures in the near future, we
may gain insights on physics at very high energy scales.

\par\vspace*{0.4cm}

\noindent {\bf Acknowledgements}\par\vspace*{0.4cm}

I am especially grateful to my collaborators Yoshiharu Kawamura and
Masaharu Yamaguchi.  I
also thank L.~Alvarez-Gaum\`e, S.~Dimopoulos, H.E.~Haber, L.J.~Hall,
C.~Kounnas, A.~Pomarol, for discussions.  Finally I thank J.~Gunion for
inviting me to this exciting workshop.
This work was supported by the Director, Office of Energy Research,
Office of High Energy and Nuclear Physics, Division of High Energy
Physics of the U.S. Department of Energy under Contract DE-AC03-76SF00098.

\end{document}